\documentclass[aps,twocolumn,amsmath,amssymb,superscriptaddress,preprintnumbers]{revtex4-1}
\usepackage[pdftex]{graphicx}
\usepackage[outdir=./]{epstopdf}
\usepackage{color}
\usepackage{enumerate,ulem}
\newcommand \beq{\begin{eqnarray}}
\newcommand \eeq{\end{eqnarray}}

\newcommand{\re}{\mathrm{Re}}
\newcommand{\im}{\mathrm{Im}}

\newcommand{\modify}[1]{{#1}}

\begin{document}

\title{Dynamical spin-to-charge conversion on the edge of quantum spin Hall insulator}
\author{Yasufumi Araki}
\affiliation{Advanced Science Research Center, Japan Atomic Energy Agency,
Tokai 319-1195, Japan}
\author{Takahiro Misawa}
\affiliation{Institute for Solid State Physics, The University of Tokyo,
 Kashiwa, Chiba 277-8581, Japan}
\author{Kentaro Nomura}
\affiliation{Institute for Materials Research, Tohoku University,
Sendai 980-8577, Japan}
\affiliation{Center for Spintronics Research Network, Tohoku University, Sendai 980-8577, Japan}

\begin{abstract}
    We theoretically manifest that the edge of a quantum spin Hall insulator (QSHI),
    attached to an insulating ferromagnet (FM),
    can realize a highly efficient spin-to-charge conversion.
    Based on a one-dimensional QSHI-FM junction,
    the electron dynamics on the QSHI edge is analyzed,
    driven by a magnetization dynamics in the FM.
    Under a large gap opening on the edge from the magnetic exchange coupling,
    we find that the spin injection into the QSHI edge gets suppressed
    while the charge current driven on the edge gets maximized,
    demanded by the band topology of the one-dimensional helical edge states.
\end{abstract}

\maketitle

\section{Introduction}

Interconversion between spin- and charge-related quantities in materials
plays an important role in manipulating spins and magnetism,
especially in the context of spintronics \cite{Dyakonov_2008,Takahashi_2008,Maekawa_spin_current}.
In particular,
the spin-charge conversion at interfaces of heterostructures has recently been studied with a great interest,
since it can make use of various novel spin-dependent properties of the electrons emergent at the interfaces \cite{Han_2018}.
The conversion phenomena at the Rashba interfaces of oxides,
the spin-momentum-locked surfaces states of topological insulators (TIs), etc.,
have been experimentally investigated \cite{Rojas-Sanchez_2013,Rojas-Sanchez_2016,Shiomi_2014,Shen_2014,Lense_2016,Wang_2016,Kondou_2016}.
\modify{The spin-to-charge conversion efficiency $\lambda_{\mathrm{sc}} \equiv - J_C^{\mathrm{(2D)}} /e J_S^{\mathrm{(3D)}} $,
defined as the ratio of
the charge current $J_C^{\mathrm{(2D)}}$ induced along the interface
to the spin current $J_S^{\mathrm{(3D)}}$ injected from the magnet via the interface,}
has been reported to reach up to a few nanometers in those systems \cite{conv-length}.


In order to improve the efficiency of the spin-to-charge conversion,
we need to reduce the spin injection $J_S$ and enhance the output current $J_C$ simultaneously.
\modify{In the present work,
we propose that a quantum spin Hall insulator (QSHI),
namely a two-dimensional (2D) TI characterized by the $\mathbb{Z}_2$ topology,
can realize a high spin-to-charge conversion efficiency $\lambda_{\mathrm{sc}}$
on its edge.}
QSHI is advantageous in spin transport in that it exhibits spin-resolved helical edge states,
which are free from backscattering by time-reversal-symmetric disorders \cite{Kane_2005,Bernevig_2006,Bernevig_2006_2}.
The spin Hall conductivity of QSHI is quantized to $e^2/h$,
which generates a quantized spin current out of an applied electric field.
This effect can be regarded as an ideal charge-to-spin conversion,
since it does not suffer from energy loss by the Joule heating.
We can thus assume that QSHI can realize the ideal spin-to-charge conversion as well.

So far it has been theoretically seen that
magnetization dynamics in a ferromagnet coupled with a QSHI
induces a charge current flowing along the junction
\cite{Qi_2008,Chen_2010,Mahfouzi_2010,Hattori_2013,Deng_2015,Wang_2019}.
From the viewpoint of the spin-to-charge conversion,
we need to understand 
\modify{how much spin should be injected to induce this edge current,
by including the effect of spin and energy dissipation from the edge.}

\begin{figure}[bp]
    \includegraphics[width=8.2cm]{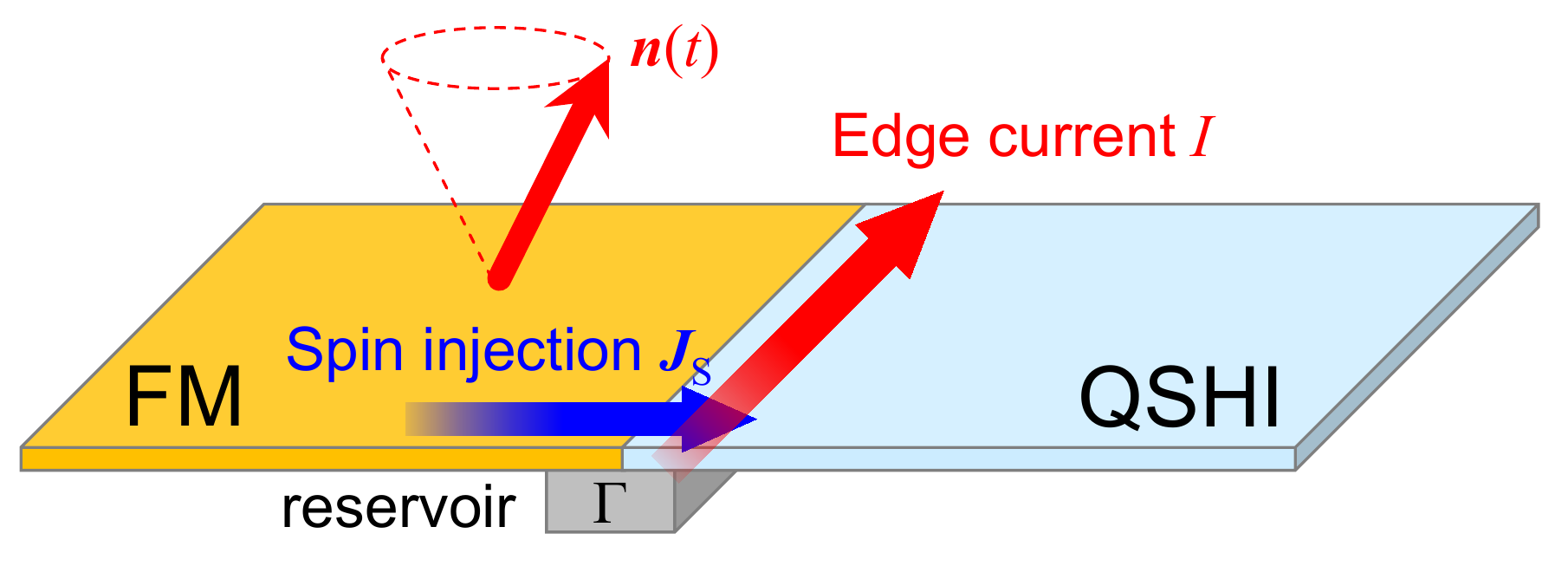}
    \caption{Schematic picture of the setup of our analysis.
    A ferromagnet (FM) and a quantum spin Hall insulator (QSHI) are coupled at their 1D boundaries.
    Phenomenologically, magnetization dynamics in the ferromagnet injects spin into the QSHI $(\boldsymbol{J}_{\mathrm{S}})$,
    which is converted into a transverse charge current on the edge of the QSHI $(I)$.
    We introduce a hypothetical metallic reservoir so that the system may maintain a periodic steady state,
    which corresponds to metallic terminals attached to the edge of the sample in experimental setups.}
    \label{fig:system-setup}
\end{figure}

In the present work,
we consider a hypothetical lateral junction of a ferromagnet and a QSHI
(see Fig.~\ref{fig:system-setup}),
to evaluate the spin-to-charge conversion efficiency of the QSHI.
Under a magnetization dynamics in the ferromagnet,
\modify{we compare the charge current $I$ induced on the edge of the QSHI,
namely the 1D counterpart of $J_C^{\mathrm{(2D)}}$,
to the spin injection rate $J_S^z$ from the ferromagnet via the QSHI edge,
namely the 2D counterpart of $J_S^{\mathrm{(3D)}}$.}
We evaluate these quantities in terms of the Floquet--Keldysh formalism \cite{Faisal_1989,Oka_2009,Tsuji_2009,Aoki_2014},
in which many-body dynamics of the electrons,
driven by the cyclic dynamics of the magnetization,
is imprinted in nonequilibrium Green's functions.

\modify{The main finding in this article is that
the QSHI edge is capable of realizing a highly efficient interfacial spin-to-charge conversion,
even in comparison with 2D interfaces including Rashba interfaces \cite{Lense_2016} and TI surfaces \cite{Wang_2016}.
Such an enhancement of $\lambda_{\mathrm{sc}}$ on the QSHI edge
stems from the insulating nature of the edge spectrum,
when it is coupled to an in-plane component of the magnetization.
Under this exchange gap,
the spin injection into the QSHI edge is reduced due to the suppression of interband transition,
whereas the current along the edge reaches its maximum value,
required by the topological pumping theory.
Under an exchange gap of $\sim 10\mathrm{meV}$
and a relaxation time typical to Dirac electron systems (e.g. graphene, TIs, etc.),
we show that $\lambda_{\mathrm{sc}}$ scales around two orders larger than those observed in Rashba interfaces and TI surfaces.
}

\modify{This article is organized as follows.
In Section \ref{sec:methods},
we define a simplified model describing a 1D edge state of QSHI coupled with a ferromagnet,
and introduce the Floquet--Keldysh formalism to treat dynamical physical quantities.
In Section \ref{sec:results},
we show our calculation results of the edge current, spin injection rate,
and their ration as the conversion efficiency $\lambda_{\mathrm{sc}}$,
and discuss when and how $\lambda_{\mathrm{sc}}$ gets enhanced on this 1D edge.
Based on these results, we give some concluding remarks in Section \ref{sec:conclusion}.
The details of our calculations are shown in the Appendices.
}

\section{Methods} \label{sec:methods}
\subsection{Model}
We start with the model of the electrons residing on the QSHI-ferromagnet junction.
The electrons on the helical edge of the QSHI,
whose spins are coupled with the magnetization $\boldsymbol{n}$ by the proximity exchange coupling $J$,
is described by the Hamiltonian
\begin{align}
    H(k) = v_\mathrm{F} k \sigma_z + J \boldsymbol{n} \cdot \boldsymbol{\sigma}.
\end{align}
in momentum space \cite{Qi_2008,Hattori_2013,Wang_2019}.
Here $v_\mathrm{F}$ is the electron Fermi velocity,
$k$ is the electron momentum along the edge,
and $\boldsymbol{\sigma}$ is the Pauli matrix for the electron spin.
If the precession of the magnetization is kept periodic around $z$-axis,
it is written as
\begin{align}
    \boldsymbol{n}(t) = (\sin\alpha \cos\Omega t, \sin\alpha \sin\Omega t, \cos\alpha), \label{eq:precession}
\end{align}
with $\alpha$ the tilting angle from $z$-axis and $\Omega$ the frequency of the precession.
Such a steady precession can be maintained, for instance,
by tuning an external magnetic field $\boldsymbol{B}_{\mathrm{ext}}$ along $z$-axis and an alternating magnetic field $\boldsymbol{B}_{\mathrm{alt}}(t)$ like a microwave,
while we shall not go into details of its mechanism.
If there is no magnetization dynamics (i.e. $\Omega = 0$),
the edge spectrum obtains a gap $2J\sin\alpha \ (\equiv 2J')$
corresponding to the in-plane component of the magnetization,
with the band dispersion $E(k) = \pm [(v_{\mathrm{F}}k + J\cos\alpha)^2 + J'^2]^{1/2}$.
The out-of-plane component $J\cos\alpha$ shifts the momentum homogeneously
and gives rise to a steady current in equilibrium,
which we shall omit in the present work.

\subsection{Floquet--Keldysh formalism}
In order to evaluate the dynamically-induced quantities carried by the electrons,
we analyze the time-periodic dynamics of the electron ensemble
in terms of the nonequilibrium Green's functions folded within a frequency domain $\Omega$,
namely the Floquet--Keldysh formalism \cite{Faisal_1989,Oka_2009,Tsuji_2009,Aoki_2014}.
The details of the analysis are left for the Appendix.
We assume that the electron dynamics reaches a so-called ``perdiodic steady state'' after a long time of driving \cite{Deghani_2014,Shirai_2016},
where the retarded/advanced/lesser Green's functions for the electrons become time-periodic,
$G^{R/A/<}(t,t') = G^{R/A/<}(t+T,t'+T)$,
with $T=2\pi/\Omega$ the precession cycle.
Using the Fourier transform within the frequency domain $\Omega$,
\begin{align}
    \mathcal{G}_{mn}(\omega) &= \int_0^T \frac{d\bar{t}}{T} \int_{-\infty}^{\infty} d\delta t \ e^{i\omega\delta t +i(m-n)\Omega \bar{t}} G(t_+,t_-), \label{eq:Keldysh-Floquet}
\end{align}
with $t_\pm \equiv \bar{t} \pm \delta t/2$,
the expectation value of the (time-independent) operator $O$ is evaluated as
\begin{align}
    \langle O(t) \rangle = -i \int_0^\Omega \frac{d\omega}{2\pi} \sum_{mn}\mathrm{Tr} \left[O \mathcal{G}^<_{mn}(\omega)\right] e^{i(m-n)\Omega t}, \label{eq:physical-quantity}
\end{align}
where the trace runs over both the momentum space and the spin space.
As the present Hamiltonian can be exactly diagonalized in the Floquet formalism,
this analysis does not require any approximations,
such as the high-frequency expansion (Floquet--Magnus expansion)
or truncation of the Floquet space,
which are commonly seen in the analyses of Floquet systems \cite{Casas_2001,Managa_2011}.
We thus evaluate the physical quantities on the edge of the QSHI,
within the whole frequency regime of the magnetization dynamics.

In order to reach a periodic steady state,
the information of the initial condition should be wiped out through dissipation to the environment.
Here we set up a hypothetical metallic reservoir coupled with the junction \cite{Han_2013},
as shown in Fig.~\ref{fig:system-setup},
which corresponds to metallic terminals in a realistic experimental setup.
Once the system reaches a periodic steady state,
the Green's functions satisfy the relation
\begin{align}
    \mathcal{G}^<(k,\omega) = \mathcal{G}^\mathrm{R}(k,\omega) \mathit{\Sigma}^<(\omega) \mathcal{G}^\mathrm{A}(k,\omega), \label{eq:lesser}
\end{align}
where the lesser self energy $\mathit{\Sigma}^<$ contains the information of electron distribution in the periodic steady state \cite{Tsuji_2009,Aoki_2014}.
While the electron distribution in driven systems in general depends on microscopic structures of the system Hamiltonian and the dissipation mechanism \cite{Deghani_2014,Shirai_2016,Iadecola_2015_2,Iadecola_2015,Seetharam_2015},
here we take a simple assumption that the lesser self energy inherits the electron distribution in the reservoir \cite{Tsuji_2009,Aoki_2014},
\begin{align}
    \mathit{\Sigma}^<_{mn}(\omega) = i\Gamma f(\omega+n\Omega) \delta_{mn}.
\end{align}
The parameter $\Gamma$ is related to the coupling between the system and the reservoir,
which is derived by integrating out the electron degrees of freedom in the reservoir \cite{Caldeira_1981,Buttiker_1985,Buttiker_1986}.
It arises as the imaginary part of the electrons' self energy,
which is encoded in the Green's functions $\mathcal{G}^{\mathrm{R/A}}$
and results in the broadening of the Floquet spectrum.
The information of the electron Fermi energy $\mu$ is also included in these Green's functions.
Here we require the temperature of the reservoir to be lower than any other energy scales in the system,
so that it can be treated as zero temperature.

\begin{figure}[tbp]
    \includegraphics[width=8.5cm]{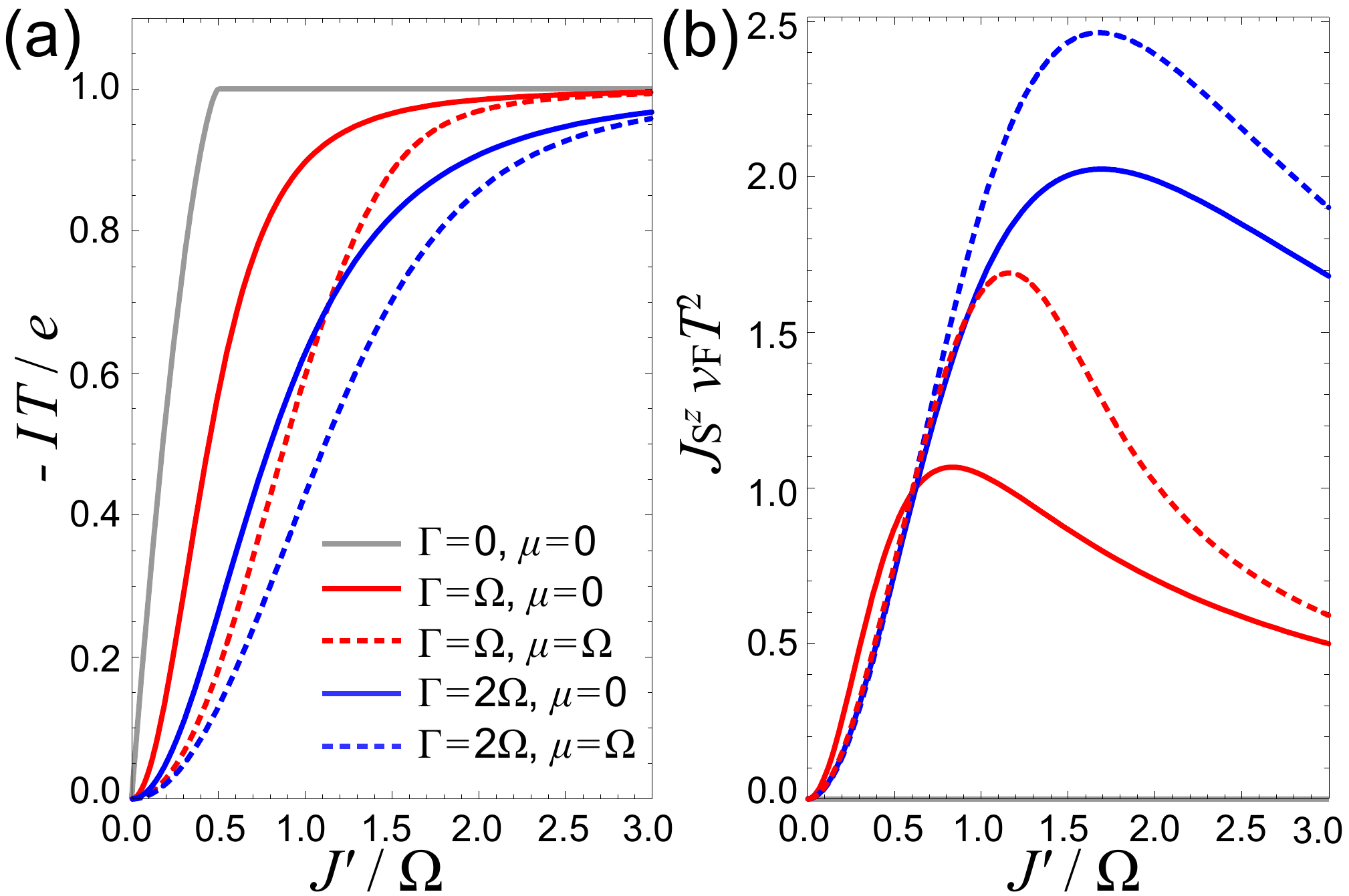}
    \caption{(a) The edge current $I$ and (b) the spin injection rate ($z$-component) $J_{\mathrm{S}}^z$ driven by the magnetization dynamics,
    parametrized by the in-plane component $J' = J \sin\alpha$ of the exchange energy at the junction.
    All the physical quantities here are rescaled by the precession frequency $\Omega$
    (or the cycle $T=2\pi/\Omega$) of the magnetization.
    The solid lines show the values obtained at charge neutrality $\mu=0$,
    while the dashed lines are obtained with $\mu$ lifted from charge neutrality.}
    \label{fig:pumped-current}
\end{figure}   


\section{Results} \label{sec:results}
\subsection{Edge current}
Let us first see the electric current driven on the 1D boundary,
which emerges as the outcome of the spin-to-charge conversion.
With the current operator
\begin{align}
    I = -e \frac{\partial H(k)}{\partial k} = -e v_\mathrm{F} \sigma_z \label{eq:current-spin-relation}
\end{align}
on the edge,
the edge current can be evaluated by Eq.~(\ref{eq:physical-quantity}),
whose behavior is shown in Fig.~\ref{fig:pumped-current}(a).
Here we
parametrize our result by $J' = J\sin\alpha$,
corresponding to the exchange gap from the in-plane component of the magnetization,
and make the physical quantities dimensionless by using the precession frequency $\Omega$.
The current $I$ is rescaled as $-I(2\pi/e\Omega) = -I T/e$,
which corresponds to the number of electrons carried per one cycle $T$.
We compare the behavior of this current
by varying the dissipation parameter $\Gamma$ and the Fermi energy $\mu$ of the electrons.

We can immediately see from this calculation result
that the edge current $I$ reaches a maximum value
\begin{align}
    I_c = -e/T \label{eq:I-C}
\end{align}
under a large exchange energy $J'$.
This behavior is obvious in the dissipationless limit $\Gamma =0$,
where $I_c$ is reached once the exchange gap $2J'$ exceeds the precession frequency $\Omega$.
The electron dynamics in this regime can be regarded ``adiabatic'',
in that an edge electron cannot be excited beyond the exchange gap.
In this regime, the induced current $I_c$ can be well described by the adiabatic pumping theory \cite{Thouless_1983},
which claims that the Berry phase accumulated by the time evolution of the electron pumps a single electron $-e$ per one cycle of precession $T$ (see Section B in the Appendix).
Within the adiabatic regime,
this quantized pumping behavior was demonstrated by various numerical and analytical schemes in previous literatures \cite{Qi_2008,Chen_2010,Mahfouzi_2010,Dora_2012}.
In the opposite regime $\Omega > 2J'$,
the magnetization dynamics can resonantly excite an edge electron from the valence band to the conduction band,
which reduces the Berry phase contribution from the valence band
and suppresses the edge current $I$ below $I_c$.
\modify{In particular, in the dissipationless limit $\Gamma=0$,
the edge current is given as
\begin{align}
    I &= -\frac{e\Omega}{2\pi} \left[ 1 - \theta(1-\delta)\left( \sqrt{1-\delta^2} - \delta \arctan\sqrt{\delta^{-2}-1} \right) \right],
\end{align}
with $\delta = 2J'/\Omega$,
which is shown by the gray line in Fig.~\ref{fig:pumped-current}(a).
}

The pumping current $I$ gets suppressed once the edge spectrum becomes metallic.
In case the Fermi level $\mu$ reaches the valence band,
the band becomes partially vacant,
leading to reduction of the Berry-phase contribution from the valence band.
If $\mu$ comes up to the conduction band, on the other hand,
it becomes partially occupied and yields the Berry-phase contribution,
which has the sign opposite to the valence-band contribution and thus partially cancels that.
Thus the current gets suppressed once the Fermi level $\mu$ is lifted from zero energy,
irrespective of its sign.
The dissipation effect $\Gamma$ by the reservoir also reduces the pumping current,
since the spectral broadening mixes up the Berry-phase contributions from the valence and conduction bands.
Such a dissipative correction to the edge current was analytically seen
in the context of photo-induced current in QSHI as well \cite{Dora_2012,Vajna_2016}.
From the above calculation results,
we can understand that the edge state needs to be insulating, with the exchange gap $2J'$, to maximize the edge current,
reaching the adiabatic pumping regime.

\subsection{Spin injection rate}
We next investigate the process of angular momentum transfer from the ferromagnet into the QSHI edge.
\modify{Whereas the spin current driven inside the electron system
is generally dependent on the spin mixing conductance under spin-orbit coupling
and requires further microscopic calculations \cite{Tserkovnyak_2002,Tserkovnyak_2005},
we here focus on the rate of angular momentum transfer, namely the loss of angular momentum, from the precessing ferromagnet,
which is straightforwardly evaluated as the counter-action of the dampinglike torque from the QSHI on the ferromagnet.}
We assume that the ferromagnet is in a thin strip geometry,
in which the constituent spins shall feel a uniform effective magnetic field from the electron spins on the QSHI edge.
If the ferromagnetic strip consists of $N$ sites per unit length,
with spin $S$ for each site,
the Landau--Lifshitz--Gilbert (LLG) equation for a single spin $\boldsymbol{S} = -S \boldsymbol{n}$ reads
\begin{align}
    \dot{\boldsymbol{S}} &= \gamma \boldsymbol{B}_{\mathrm{eff}} \times \boldsymbol{S} -\alpha_{\mathrm{d}} \boldsymbol{S} \times \dot{\boldsymbol{S}} /S,
\end{align}
where $\gamma = g \mu_{\mathrm{B}}$ denotes the gyromagnetic ratio
and $\alpha_{\mathrm{d}}$ is the Gilbert damping parameter intrinsic to the ferromagnet.
The effective magnetic field $\boldsymbol{B}_{\mathrm{eff}}$ for a single spin is given as $\boldsymbol{B}_{\mathrm{eff}} = \boldsymbol{B}_{\mathrm{ext}} + \boldsymbol{B}_{\mathrm{alt}} + \boldsymbol{B}_{\mathrm{el}}$,
where $\boldsymbol{B}_{\mathrm{el}} \equiv -J\langle \boldsymbol{\sigma} \rangle / \gamma N S$ is the contribution from the electron spin accumulation $\langle \boldsymbol{\sigma} \rangle$ on the QSHI edge.
The torque from the effective field $\boldsymbol{B}_{\mathrm{el}}$ is
\begin{align}
    \boldsymbol{t}_{\mathrm{el}} = \gamma \boldsymbol{B}_{\mathrm{el}} \times \boldsymbol{S} = (J/N) \langle \boldsymbol{\sigma} \rangle \times \boldsymbol{n},
\end{align}
which arises as the feedback effect from the edge electrons onto the ferromagnet.

Let us here consider the net feedback torque on the spins within a unit length of the strip,
given by $\boldsymbol{T}_{\mathrm{el}} = N \boldsymbol{t}_{\mathrm{el}} = J \langle \boldsymbol{\sigma} \rangle \times \boldsymbol{n}$.
This torque can be separated into the field-like component $\boldsymbol{T}^{\mathrm{f}} \propto \boldsymbol{e}_z \times \boldsymbol{n}$
and the damping-like component $\boldsymbol{T}^{\mathrm{d}} \propto \dot{\boldsymbol{n}} \times \boldsymbol{n}$.
Whereas the field-like component $\boldsymbol{T}^{\mathrm{f}}$ gives a correction to the external magnetic field $\boldsymbol{B}_{\mathrm{ext}} \parallel z$
that maintains the spin precession around $z$-axis,
the damping-like component $\boldsymbol{T}^{\mathrm{d}}$ yields a correction to the Gilbert damping parameter $\alpha_{\mathrm{d}}$.
$\boldsymbol{T}^{\mathrm{d}}$ gives a negative angular momentum transfer
from the conduction electrons to the ferromagnet,
which \modify{is the counter-action of} the spin injection from the ferromagnet into the conduction electrons \cite{Silsbee_1979,Tserkovnyak_2002}.
Therefore, in order to understand the spin injection behavior,
we need to evaluate the damping-like torque $\boldsymbol{T}^{\mathrm{d}}$.

Among the three components of the electron spin accumulation $\langle \sigma_{x,y,z}(t)\rangle$,
the damping-like torque $\boldsymbol{T}^{\mathrm{d}} \propto \dot{\boldsymbol{n}} \times \boldsymbol{n}$ comes from the component parallel to $\dot{\boldsymbol{n}}(t) = \Omega \sin\alpha(-\boldsymbol{e}_x \sin\Omega t + \boldsymbol{e}_y \cos\Omega t)$.
By denoting this component in $\langle \boldsymbol{\sigma}(t)\rangle$ as $\boldsymbol{\sigma}_{\mathrm{d}}(t) = \sigma_{\mathrm{d}}(-\boldsymbol{e}_x \sin\Omega t + \boldsymbol{e}_y \cos\Omega t)$,
the time average of $\boldsymbol{T}^{\mathrm{d}}$ is given as
\begin{align}
    \overline{\boldsymbol{T}}^{\mathrm{d}} &= \int_0^T \frac{dt}{T} [J \boldsymbol{\sigma}_{\mathrm{d}}(t) \times \boldsymbol{n}(t) ] = -J \sigma_{\mathrm{d}} \sin\alpha \ \boldsymbol{e}_z,
\end{align}
from which we obtain the spin angular momentum
\begin{align}
    \boldsymbol{J}_{\mathrm{S}} = -\overline{\boldsymbol{T}}^{\mathrm{d}} = J' \sigma_{\mathrm{d}} \boldsymbol{e}_z
\end{align}
transferred from the ferromagnet to the QSHI edge,
per unit time and unit length in average.
We can thus straightforwardly calculate the spin injection rate $\boldsymbol{J}_{\mathrm{S}}$,
by evaluating the spin accumulation $\langle \boldsymbol{\sigma}(t)\rangle$
based on the Floquet--Keldysh formalism (see the Appendix for details).

The behavior of $J_{\mathrm{S}}^z$ parametrized by $J' = J \sin\alpha$ is shown in Fig.~\ref{fig:pumped-current} (b).
Here $J_{\mathrm{S}}^z$, having the dimension of $[\mathrm{time}]^{-1} [\mathrm{length}]^{-1}$,
is rescaled by multiplying the time scale $T$ and the length scale $v_{\mathrm{F}} T$.
If the system is isolated from the environment,
corresponding to the dissipationless limit $\Gamma \rightarrow 0$,
$J_{\mathrm{S}}^z$ vanishes:
since the edge electrons do not lose spin angular momentum in this limit,
the periodic steady state is maintained without injecting spin continuously.
The spin injection $J_{\mathrm{S}}^z$ arises due to the loss of spin in the reservior.
We should note here that the spin injection is suppressed in the ``adiabatic'' regime $J' \gtrsim \Omega/2, \mu, \Gamma$,
with its asymptotic behavior
\begin{align}
    J_{\mathrm{S}}^z \overset{J' \rightarrow \infty}{\approx} \frac{\Omega \Gamma^2}{8\pi v_{\mathrm{F}} J'}. \label{eq:asym-JS}
\end{align}
Since the edge electron can hardly be excited
beyond the exchange gap in this regime,
we can understand that the spin injection process,
accompanied with a spin filp of the edge electron, is suppressed.
On the other hand,
in case the Fermi energy $\mu$ or the spectral broadening $\Gamma$ exceeds the exchange energy $J'$,
the system becomes metallic and thus admits a large spin injection.

Spin torque is accompanied with energy transfer as well.
The damping-like torque by the effective field $\boldsymbol{B}_{\mathrm{eff}}$
exerts a negative work on the spins,
with its power $-p = -\gamma \boldsymbol{B}_{\mathrm{eff}} \cdot \dot{\boldsymbol{S}}$ on a single spin \cite{Maekawa_spin_current,Tserkovnyak_2005};
for the spins within a unit length, its power is given as
\begin{align}
    -P = -Np = -J \langle \boldsymbol{\sigma} \rangle \cdot \dot{\boldsymbol{n}} = -\Omega J \sigma_{\mathrm{d}} \sin\alpha.
\end{align}
Therefore,
we can see that energy $P = \Omega J' \sigma_{\mathrm{d}}$ is injected from the ferromagnet to the edge electrons of the QSHI per unit time and unit length.
The spin injection rate $J_{\mathrm{S}}^z$ and the energy injection rate $P$
satisfy a simple relation $P = \Omega J_{\mathrm{S}}^z$.
This relation can be attributed to the magnon exchange picture:
if we consider the constituent spins in the ferromagnet as quantum spins,
their precession modes can be quantized as magnons,
where the uniform (Kittel) mode carries spin $1$ and energy $\Omega$.
In this picture, the spin injection can be regarded as
the flow of magnons from the ferromagnet into the QSHI edge,
which requires the proportionality between the injected spin $J_{\mathrm{S}}^z$ and energy $P$.
Once we attach a reservior, or terminals, to the system to extract the transport properties,
there arises a loss of spin and energy in the reservoir,
leading to a continious injection of spin $J_{\mathrm{S}}^z$ and energy $P$
that maintains the periodic steady state.

\begin{figure}[tbp]
    \includegraphics[width=8.5cm]{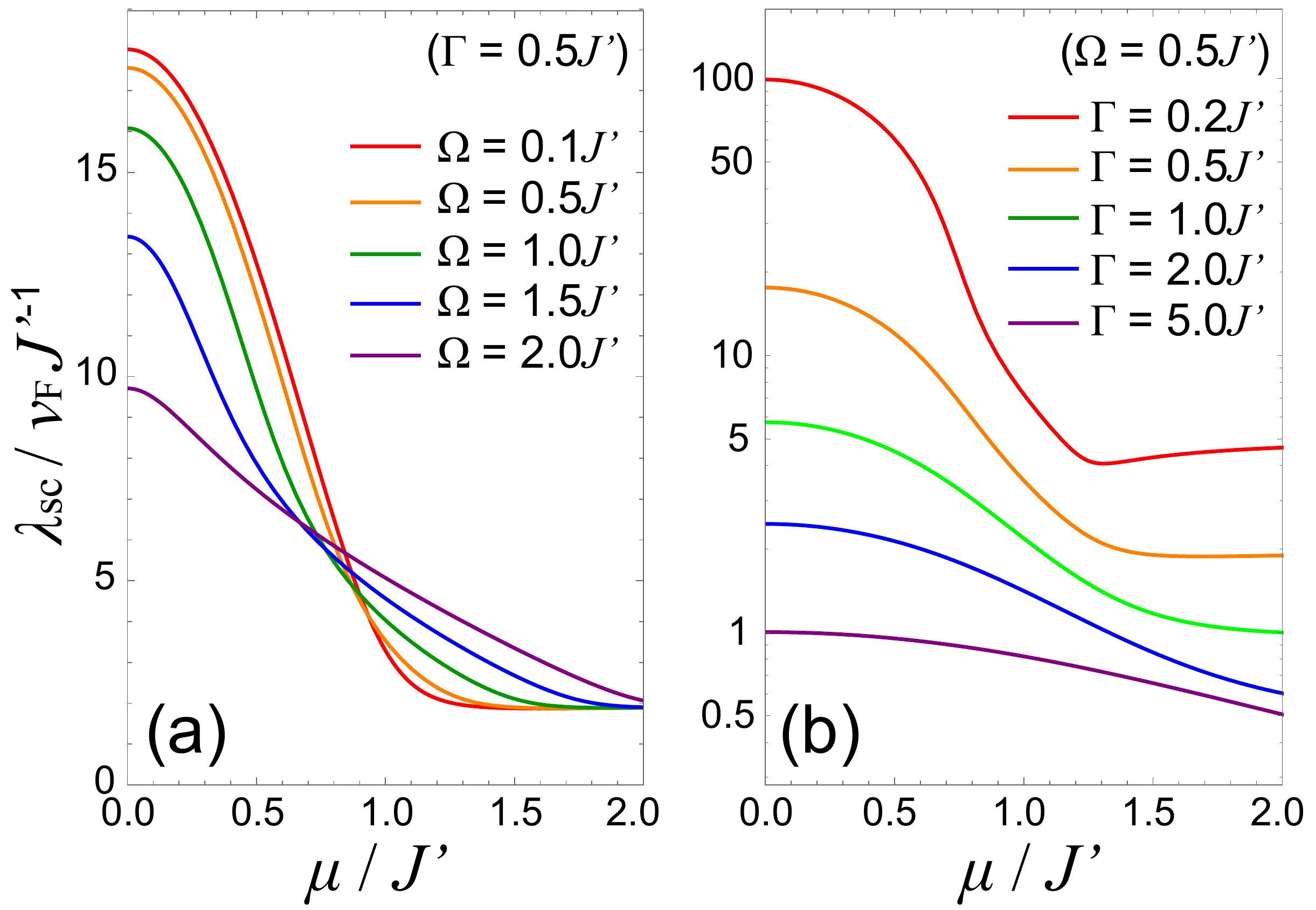}
    \caption{The spin-to-charge conversion efficiency $\lambda_{\mathrm{sc}} \equiv -I/e J_{\mathrm{S}}^z$ parametrized by the Fermi level $\mu$.
    All the physical quantities here are rescaled by fixing the exchange gap parameter $J' = J \sin\alpha$.
    Panel (a) shows $\lambda_{\mathrm{sc}}$ by varying the precession frequency $\Omega$ of the magnetization,
    while (b) is the logarithmic plot of $\lambda_{\mathrm{sc}}$ by varying the dissipation parameter $\Gamma$ of the reservoir.
    From these calculation results, one can see that $\lambda_{\mathrm{sc}}$ is highly enhanced when $J'$ is dominant over the other energy scales, $\mu$, $\Omega$, and $\Gamma$.
    }
    \label{fig:conversion-efficiency}
\end{figure}   

\subsection{Spin-to-charge conversion efficiency}
Finally,
we evaluate the efficiency of the spin-to-charge conversion on the 1D edge.
The conversion efficiency
\modify{$\lambda_{\mathrm{sc}}$ is defined as the ratio of the induced edge current $I$
to the spin injection rate $J_{\mathrm{S}}^z$,
namely the loss of spin angular momentum from the ferromagnet,}
\begin{align}
    \lambda_{\mathrm{sc}} \equiv -\frac{I/e}{J_{\mathrm{S}}^z},
\end{align}
\modify{which characterizes how much charge current the system can generate by consuming spin angular momentum in the ferromagnet.}
This ratio has the dimension of length,
which is the same as that defined at 2D interfaces.
It works for the energy efficiency for inducing the edge current, namely $I/P$,
as well,
due to the proportionality $P = \Omega J_{\mathrm{S}}^z$ stated above.
By fixing the exchange energy $J'$ and varying the Fermi energy $\mu$ of the electrons,
the spin-to-charge conversion rate $\lambda_{\mathrm{sc}}$ on the QSHI edge is obtained as shown in Fig.~\ref{fig:conversion-efficiency}.
We can see from this calculation result
that the conversion on the edge becomes highly efficient if the Fermi level $\mu$ is deeply inside the exchange gap, i.e. $|\mu| \ll J'$,
since the current $I$ reaches the constant value $-e/T$ demanded by the adiabatic pumping theory,
whereas the spin injection $J_{\mathrm{S}}^z$ gets suppressed by the exchange gap.
This behavior becomes significant if the magnetization dynamics is adiabatic,
i.e. $\Omega < 2J'$,
as shown in Fig.~\ref{fig:conversion-efficiency}(a),
so that it may not excite an electron beyond the exchange gap.
We also need a low dissipation effect $\Gamma$ by the reservoir (terminals)
to achieve the highly efficient spin-to-charge conversion,
as seen from Fig.~\ref{fig:conversion-efficiency}(b).

\modify{As we can see from Fig.~\ref{fig:pumped-current},
the edge current $I$ is not directly proportional to the spin injection rate $J_{\mathrm{S}}^z$;
in particular, in the dissipationless limit $\Gamma =0$,
the current $I$ is driven even though the spin injection rate $J_{\mathrm{S}}^z=0$.
This relation implies that the spin-to-charge conversion behavior here cannot be interpreted as the inverse spin Hall effect,
where the injected spin current is converted to a charge current either intrinsically or extrinsically \cite{Averkiev_1983,Sinova_2015,Saitoh_2006,Valenzuela_2006}.
We can rather understand that
this effect is similar to the inverse Edelstein effect (IEE) observed at 2D interfaces,
where the spin accumulation at the interface induced by the magnetization dynamics
is the origin of the interfacial charge current \cite{Shen_2014,Edelstein_1990,Ganichev_2002}.
This mechanism is justified by the operator relation Eq.~(\ref{eq:current-spin-relation}),
between the current $I$ and the electron spin $\sigma_z$.
Under a finite dissipation effect $\Gamma$,
the spin accumulation $\langle \sigma_z \rangle$ is subject to relaxation.
By denoting the spin relaxation time as $\tau_s$,
the spin relaxation is balanced with the spin injection $J_{\mathrm{S}}^z$ in the steady state as
\begin{align}
    J_{\mathrm{S}}^z - \frac{\langle \sigma_z \rangle}{\tau_s} =0.
\end{align}
We can therefore relate $J_{\mathrm{S}}^z$ and $I$ phenomenologically as
\begin{align}
    I = -e v_{\mathrm{F}} \langle \sigma_z \rangle = -e v_{\mathrm{F}} \tau_s J_{\mathrm{S}}^z,
\end{align}
yielding $\lambda_{\mathrm{sc}} = v_{\mathrm{F}} \tau_s$,
which is quite similar to the case of IEE at the 2D surface of TI
(see Refs.~\onlinecite{Rojas-Sanchez_2016}, \onlinecite{Shiomi_2014}, and their Supplemental Materials).
}

\modify{The main difference between
the present spin-to-charge conversion effect on the 1D edge of QSHI
and the IEE at 2D interfaces is the emergence of an exchange gap in the electron system.
While the electrons are inevitably subject to spin relaxation and Joule heating
due to the scattering by impurities at metallic interfaces,
the current driven on the edge of QSHI is nearly free from dissipation
as long as the Fermi level is inside the exchange gap.
This implies that the spin relaxation time $\tau_s$ is largely dependent on the system parameters $J', \mu, \Gamma$, and $\Omega$,
leading to the variation in the conversion efficiency $\lambda_{\mathrm{sc}}$ shown in Fig.~\ref{fig:conversion-efficiency}.
}

\modify{Under a large exchange gap $J'$,
the spin relaxation gets strongly suppressed,
which enables us to drive the electric current without consuming spin angular momentum from the ferromagnet,
i.e. $J_{\mathrm{S}}^z \rightarrow 0$.
Using the asymptotic behavior of $J_{\mathrm{S}}^z$ shown by Eq.~(\ref{eq:asym-JS}),}
the asymptotic behavior of $\lambda_{\mathrm{sc}}$ in this regime is given as
\begin{align}
    \lambda_{\mathrm{sc}} \overset{J' \rightarrow \infty}{\approx}  4 v_{\mathrm{F}}J' \Gamma^{-2}. \label{eq:conv-asymptotic}
\end{align}
\modify{By using the typical scales calculated and observed in graphene,
$v_{\mathrm{F}} \approx 10^{5}\mathrm{m/s}$ (Refs.~\onlinecite{Novoselov_2005,CastroNeto_2009}),
$\Gamma \approx 10 \mathrm{meV}$ (from the single-particle level broadening estimated in Refs.~\onlinecite{Hwang_2008,Hong_2009}),
and $J'\approx 10\mathrm{meV}$ (Refs.~\onlinecite{Usachov_2015,Rybkin_2018}),
we can roughly estimate $\lambda_{\mathrm{sc}} \approx 10^2 \mathrm{nm}$.}
\modify{For instance, if we desire an output current $I = -1 \mathrm{nA}$,
the precession frequency $\Omega = 39 \mathrm{GHz}$ is required from Eq.~(\ref{eq:I-C}),
which gives the spin injection rate $J_{\mathrm{S}}^z = 6 \mathrm{mA}/(e\cdot\mathrm{m})$ from the asymptotic form Eq.~(\ref{eq:asym-JS}).
In this case, the conversion efficiency $\lambda_{\mathrm{sc}}$ reaches around $170 \mathrm{nm}$.}
\modify{Compared to $\lambda_{\mathrm{sc}} \lesssim 6 \mathrm{nm}$ at the 2D Rashba interfaces of complex oxides \cite{Lense_2016}
and $\lambda_{\mathrm{sc}} \lesssim 0.04 \mathrm{nm}$ at the surfaces of TIs \cite{Wang_2016}
observed in the experiments,
we expect that the 1D edge of QSHI can realize the spin-to-charge conversion efficiency
around two orders greater than those reported in 2D interfaces.}

\section{Conclusion} \label{sec:conclusion}
In this article,
we have theoretically investigated the dynamical spin-to-charge conversion phenomenon
on the edge of a 2D QSHI.
By taking a hypothetical lateral junction of a 2D ferromagnet and a QSHI,
we have evaluated the spin-to-charge conversion efficiency on the edge of the QSHI,
driven by magnetization dynamics in the ferromagnet.
The main finding in this article is that the conversion efficiency
is highly enhanced,
under a large exchange gap on the edge spectrum induced by the in-plane component of the magnetization.
In contrast to the conventional spin pumping phenomena in metals,
the edge state should be insulating to idealize the spin-to-charge conversion,
since the converted charge current is based on the topological origin,
namely the adiabatic charge pumping.
The electrons on the 1D helical edge states is completely free from scattering by charged disorders,
which minimizes the leakage of spin and energy in this spin--to-charge conversion process
as long as the coupling to the terminals or environment is weak enough.

In order to make the best of the topological characteristics of the QSHI edge
for realizing the ideal spin-to-charge conversion,
we find the following criteria from our calculations:
(i) the Fermi level $\mu$ should lie inside the exchange gap $(|\mu| < J')$,
(ii) the precession frequency $\Omega$ of the magnetization should not exceed the exchange gap $(\Omega <2J')$,
and (iii) the exchange gap should be well resolved against the spectral broadening $(\Gamma < J')$.

Our findings imply that a 2D QSHI can serve as an efficient detector of a spin current,
nearly free from the leakage of spin and energy.
Using layered QSHI materials,
such as the transition metal dichalcogenide $1T'-\mathrm{WTe_2}$ \cite{Qian_2014,Zheng_2016,Tang_2017},
monolayer germanene (Ge) or stanene (Sn) \cite{Liu_2011,Xu_2013,Molle_2017,Deng_2018}
reported in recent studies,
one can expect a flexible design of highly integrated spin-charge devices.
Thin films of topological Dirac semimetals (e.g. $\mathrm{Cd_3 As_2}, \mathrm{Na_3 Bi}$),
characterized by the $\mathbb{Z}_2$ topology,
are also seen to exhibit the QSHI phase \cite{Niu_2017,Collins_2018}.
It has been numerically simulated that the topological Dirac semimetals
also exhibit the dynamical spin-to-charge conversion,
carried by the spin-resolved Fermi-arc surface states \cite{Misawa_2019}.

\modify{In the present analysis,
we have taken into account the fermionic reservoir,
corresponding to the metallic leads attached to the system,
as the main source of the dissipation.
In realistic dynamical systems, phonons contribute to relaxation as well,
whose effect on electron distribution has been intensely studied
in the context of the Floquet dynamics \cite{Deghani_2014,Iadecola_2015_2,Iadecola_2015,Seetharam_2015}.
We can qualitatively expect that, even if the reservoir is bosonic,
the spin injection is suppressed and the edge current is quantized under a sufficiently large exchange gap,
since the interband excitation by a magnon absorption is almost forbidden by the exchange gap.
Under a small exchange gap, the conduction band of the edge may exhibit a finite occupation probability, leading to reduction of the conversion efficiency $\lambda_{\mathrm{sc}}$.
The detailed behavior of the reduction of $\lambda_{\mathrm{sc}}$
within such a non-adiabatic regime requires one
to solve the time evolution of the electron distribution explicitly under the system-reservoir coupling,
which is left for a further analysis.}

\modify{Since the induced current circulates along the edge of the QSHI,
the present spin-to-charge conversion phenomenon can also be regarded
as the conversion from the injected spin into an orbital magnetization of the QSHI,
although the orbital magnetization is not directly evaluated in this work.
While it is known that QSHIs and topological semimetals in equilibrium show the crossed correlation between spin and orbital magnetizations
\cite{Nakai_2016,Ominato_2019},
the present result implies its nonequilibrium counterpart,
which is also left for our theoretical interest.}

\acknowledgments{
Y.~A. is supported by JSPS KAKENHI Grant Number JP17K14316
and the Leading Initiative for Excellent Young Researchers (LEADER).
T.~M. is supported by JSPS KAKENHI Grants Numbers JP16H06345, JP16K17746, JP19K03739, 
and by Building of Consortia for the Development of Human Resources in 
Science and Technology from the MEXT of Japan.
K.~N. is supported by JSPS KAKENHI Grant Numbers JP15H05854 and JP17K05485,
and JST CREST Grant No.~JPMJCR18T2.
}

\appendix

\section{Keldysh--Floquet treatment of nonequilibrium physical quantities}
In this part,
we give a detailed explanation on our analytical treatment
of the dynamics of the ensemble of edge electrons,
based on the Keldysh--Floquet formalism.

\subsection{One-particle states in Floquet picture}
Let us first start with the dynamics of one-particle state.
Since the edge Hamiltonian under this precession is time-periodic,
i.e. $H(k,t) = H(k,t+T)$ (with the periodicity $T=2\pi/\Omega$),
the electron dynamics can be treated in terms of the Floquet theory.
The time dependence in the solution of the Schr\"{o}dinger equation $H(k,t)|\Psi_\alpha(k,t)\rangle = i\partial_t |\Psi_\alpha(k,t)\rangle$ is expanded as
\begin{align}
    |\Psi_\alpha(k,t)\rangle = e^{-i\mathcal{E}_\alpha(k)t} \sum_{n \in \mathbb{Z}} e^{-in\Omega t} |\phi_\alpha^n(k)\rangle,
\end{align}
where its quasienergy $\mathcal{E}_\alpha(k)$ and the expanded components $|\phi_\alpha^n(k)\rangle$ are related by the Floquet equation
\begin{align}
    \sum_n \mathcal{H}_{mn}(k) |\phi_\alpha^n(k)\rangle = \mathcal{E}_\alpha(k) |\phi_\alpha^m(k)\rangle, \label{eq:Floquet-Schrodinger}
\end{align}
with the ``Floquet Hamiltonian''
\begin{align}
    \mathcal{H}_{mn}(k) = \frac{1}{T} \int_0^T dt \ e^{i(m-n)\Omega t} H(k,t) - n\Omega \delta_{mn}. \label{eq:Floquet-Hamiltonian-def}
\end{align}
(It should be noted that some literatures use the terminology ``Floquet Hamiltonian'' for $\tilde{\mathcal{H}}_{mn} = \mathcal{H}_{mn} + n\Omega\delta_{mn}$.)
Thus the time-dependent solution $|\Psi_\alpha(k,t)\rangle$ based on the original Hilbert space $\mathbb{H}$
is mapped to the time-independent wave function $|\Phi_\alpha(k)\rangle\!\rangle \equiv \{ |\phi_\alpha^n(k)\rangle \}_{n \in \mathbb{Z}}$
based on the ``extended'' Hilbert space $\mathbb{H} \times \mathbb{Z}$,
which is required to satisfy the infinite-dimensional eigenvalue equation $\mathcal{H} |\Phi_\alpha\rangle\!\rangle = \mathcal{E}_\alpha |\Phi_\alpha\rangle\!\rangle$.
The Floquet Hamiltonian $\mathcal{H}$ here reads
\begin{align}
    \mathcal{H}_{mn}(k) = \begin{pmatrix} [\epsilon_0(k)-m\Omega] \delta_{mn} & J' \delta_{m,n+1} \\ J' \delta_{m,n-1} & -[\epsilon_0(k)+m\Omega] \delta_{mn} \end{pmatrix}, \label{eq:Floquet-Hamiltonian}
\end{align}
with $\epsilon_0(k) = v_\mathrm{F} k + J\cos\alpha, \ J' = J \sin\alpha$.

\begin{figure}[tbp]
    \includegraphics[width=8.4cm]{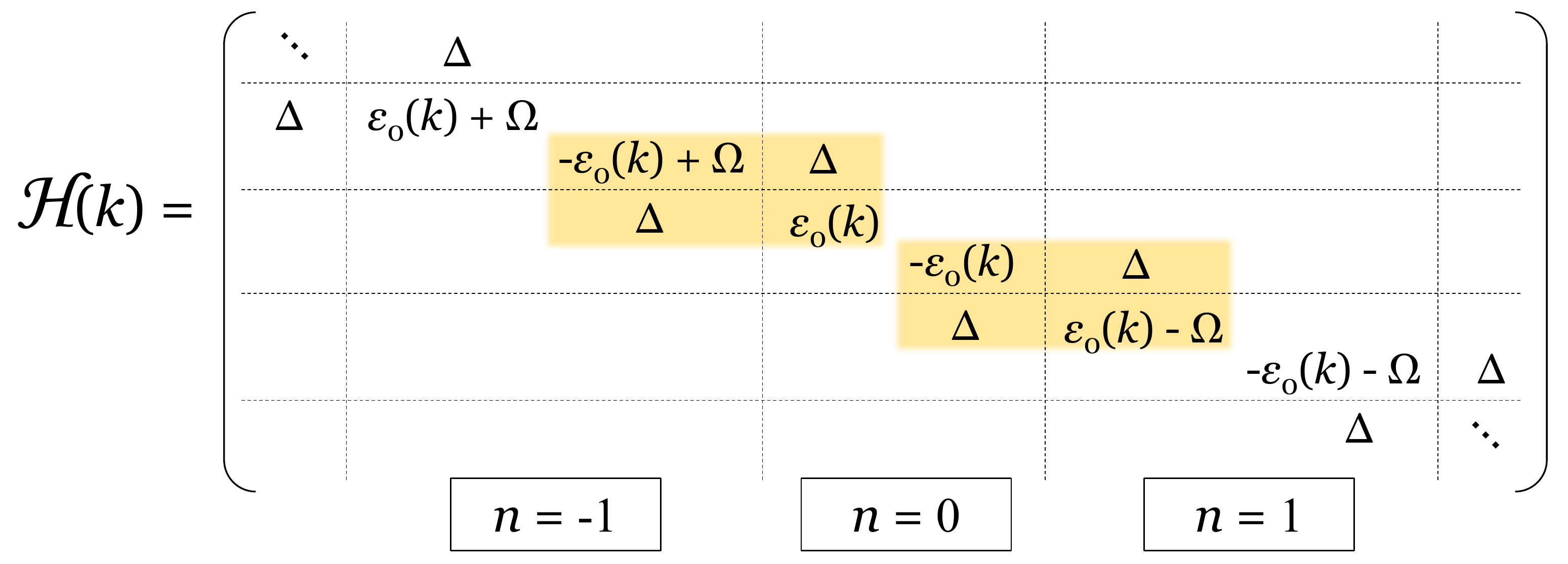}
    \caption{Matrix structure of the Floquet Hamiltonian $\mathcal{H}(k)$.
    The matrix becomes block diagonal,
    with each block spanned by $\{ |\nu-\tfrac{1}{2},\downarrow\rangle , |\nu+\tfrac{1}{2},\uparrow\rangle \}_{\nu \in \mathbb{Z}+1/2}$.}
    \label{fig:matrix-structure}
\end{figure}   

The Floquet index $n$,
related with the phase factor $e^{-in\Omega t}$ in the time-dependent solution,
accounts for the number of ``energy quanta'' of $\Omega$
arising from the time-periodic dynamics in the system;
here the energy quantum is a magnon (with spin 1) arising from the precession of magnetization.
The off-diagonal components in the Floquet Hamiltonian [Eq.~(\ref{eq:Floquet-Hamiltonian})] couples the neighboring magnon number sectors.
The top-right component in Eq.~(\ref{eq:Floquet-Hamiltonian}) adds one magnon and flips the electron spin from $\downarrow$ to $\uparrow$,
and vice versa for the bottom-left component.
Thus the Floquet matrix $\mathcal{H}$ becomes block diagonal:
each subspace $\mathbb{H}_\nu (\nu \in \mathbb{Z}+\tfrac{1}{2})$
spanned by the basis
$\{ |n= \nu-\tfrac{1}{2};\sigma_z = \downarrow\rangle,|n= \nu+\tfrac{1}{2}; \sigma_z = \uparrow\rangle \}$ gets decoupled from one another (see Fig.~\ref{fig:matrix-structure}).
The block in this subspace $\mathbb{H}_\nu$ reads
\begin{align}
\mathcal{H}_\nu(k) &= -\nu\Omega + \begin{pmatrix} -\epsilon_0(k) +\tfrac{\Omega}{2} & J' \\ J' & \epsilon_0(k)-\tfrac{\Omega}{2} \end{pmatrix},
\end{align}
where the upper and lower components correspond to $|\nu-\tfrac{1}{2},\downarrow\rangle $ and $|\nu+\tfrac{1}{2},\uparrow\rangle$, respectively.
Omitting the constant shift of the energy,
$\mathcal{H}_\nu(k)$ is exactly the same as the equilibrium edge Hamiltonian under a Zeeman field,
\begin{align}
    \bar{\mathcal{H}}(k) &= \begin{pmatrix} -\epsilon_0(k) +\tfrac{\Omega}{2} & J' \\ J' & \epsilon_0(k)-\tfrac{\Omega}{2} \end{pmatrix}.
\end{align}
Therefore, the Floquet Hamiltonian $\mathcal{H}(k)$ can be exactly diagonalized,
yielding the quasienergies
\begin{align}
    \mathcal{E}_{\nu \pm} (k) = -\nu\Omega \pm \sqrt{\left[\epsilon_0(k) - \tfrac{\Omega}{2}\right]^2 +J'^2}. \label{eq:quasienergy}
\end{align}
We denote the corresponding eigenstates as $|\Phi_\nu^\pm(k) \rangle\!\rangle$,
which are based on the subspace $\mathbb{H}$.
Due to the redundancy of the extended Hilbert space,
it is just enough to focus on one of the subspaces $\{ \mathbb{H}_\nu \}$.
The helical edge states with $|-1,\downarrow\rangle$ and $|0,\uparrow\rangle$ get hybridized by the magnon exchange,
leading to gap opening $2J'$ at $\epsilon_0(k) = \tfrac{\Omega}{2}$,
i.e. $k = k_0 \equiv v_\mathrm{F}^{-1} \left(\tfrac{\Omega}{2}-J\cos\alpha\right)$.
It should be noted that the present Floquet Hamiltonian can be thus exactly diagonalized regardless of the precession periodicity,
in contrast to the adiabaticity (low frequency) needed for the Thouless pumping theory,
or the Magnus expansion (high frequency expansion) applied to ordinary Floquet systems.

\subsection{Keldysh--Floquet treatment}
We now move on to the many-body dynamics.
The many-body dynamics in the periodic steady state
is described by the nonequilibirum Green's functions
based on the extended Hilbert space $\mathbb{H} \times \mathbb{Z}$,
defined by Eq.~(3).
If the periodic steady state is ensured by the dissipation
into the metallic reservior (fermionic heat bath),
the system inherits the electron distribution in the reservior,
and the lesser Green's function satisfies Eq.~(5),
\begin{align}
    \mathcal{G}^<(k,\omega) = \mathcal{G}^\mathrm{R}(k,\omega) \mathit{\Sigma}^<(\omega) \mathcal{G}^\mathrm{A}(k,\omega),
\end{align}
with
$\mathit{\Sigma}^<_{mn}(\omega) = i\Gamma f(\omega+n\Omega) \delta_{mn}$.
The retarded and advanced Green's functions for the dissipative time evolution are given by
\begin{align}
    \mathcal{G}^{\mathrm{R/A}}(k,\omega) = \left[\omega +\mu -\mathcal{H}(k) \pm i\tfrac{\Gamma}{2}\right]^{-1},
\end{align}
with the chemical potential $\mu$.
Here the parameter $\Gamma$ can be microscopically obtained
by integrating out the electron degrees of freedom in the reservoir;
we regard $\Gamma$ as a phenomenological parameter in our analysis here.

In the present case, since the Floquet Hamiltonian $\mathcal{H}(k)$ is block diagonal 
within each subspace $\mathbb{H}_\nu$ spanned by $\{|\nu-\tfrac{1}{2}\downarrow\rangle,|\nu+\tfrac{1}{2},\uparrow\rangle\}$,
the Green's functions $\mathcal{G}^{\mathrm{R}/\mathrm{A}/<}$ are also block diagonal.
The Green's functions in the subspace $\mathbb{H}_\nu$ read
\begin{align}
    \mathcal{G}^{\mathrm{R}/\mathrm{A}}_\nu(k,\omega) &= \left[\omega+\mu+\nu\Omega - \bar{\mathcal{H}}(k) \pm i\tfrac{\Gamma}{2} \right]^{-1} \\
    \mathcal{G}^{<}_\nu(k,\omega) &= i\Gamma \mathcal{G}^{\mathrm{R}}_\nu(k,\omega) \mathcal{F}_\nu(\omega) \mathcal{G}^{\mathrm{A}}_\nu(k,\omega),
\end{align}
with
\begin{align}
    \mathcal{F}_\nu(\omega) \equiv \mathrm{diag} \left\{ f(\omega+(\nu-\tfrac{1}{2})\Omega), \ f(\omega+(\nu+\tfrac{1}{2})\Omega) \right\}.
\end{align}
We should note here that, similarly to the Floquet quasienergies $\mathcal{E}(k)$,
these Green's functions also show redundancy in energy:
they can be written with the $2 \times 2$ matrices $\bar{\mathcal{G}}^{\mathrm{R}/\mathrm{A}/<}(k,\bar{\omega})$ as
\begin{align}
    \mathcal{G}^{\mathrm{R}/\mathrm{A}/<}_\nu(k,\omega) = \bar{\mathcal{G}}^{\mathrm{R}/\mathrm{A}/<} \left(k,\omega+\nu\Omega \right),
\end{align}
where
\begin{align}
    \bar{\mathcal{G}}^{\mathrm{R}/\mathrm{A}}(k,\bar{\omega}) &= \left[\bar{\omega} +\mu - \bar{\mathcal{H}}(k) \pm i\tfrac{\Gamma}{2} \right]^{-1} \\
    \bar{\mathcal{G}}^{<}(k,\bar{\omega}) &= i\Gamma \bar{\mathcal{G}}^{\mathrm{R}}(k,\bar{\omega}) \bar{\mathcal{F}}(\bar{\omega}) \bar{\mathcal{G}}^{\mathrm{A}}(k,\bar{\omega}) \\
    \bar{\mathcal{F}}(\bar{\omega}) &= \mathrm{diag} \left\{ f(\bar{\omega}-\tfrac{\Omega}{2}), \ f(\bar{\omega}+\tfrac{\Omega}{2}) \right\}.
\end{align}
In particular,
each component in these Green's functions is given as
\begin{align}
    \bar{\mathcal{G}}^{\mathrm{R/A}} &= 
    \frac{1}{D^{\mathrm{R/A}}_\downarrow D^{\mathrm{R/A}}_\uparrow - J'^2}
    \begin{pmatrix}
        D^{\mathrm{R/A}}_{\uparrow} & J' \\ J' & D^{\mathrm{R/A}}_{\downarrow}
    \end{pmatrix} \\
    \bar{\mathcal{G}}^< &= 
    \frac{i\Gamma}{\left[ D^{\mathrm{R}}_{\downarrow} D^{\mathrm{R}}_{\uparrow} - J'^2 \right] \left[ D^{\mathrm{A}}_{\downarrow} D^{\mathrm{A}}_{\uparrow} - J'^2 \right]} \label{eq:G-lesser-matrix} \\
    & \quad \times \begin{pmatrix}
        f_\downarrow D^{\mathrm{R}}_\uparrow D^{\mathrm{A}}_\uparrow + f_\uparrow J'^2 &
        f_\downarrow J' D^{\mathrm{R}}_\uparrow + f_\uparrow J'D^{\mathrm{A}}_\downarrow \\
        f_\downarrow J' D^{\mathrm{A}}_\uparrow + f_\uparrow J'D^{\mathrm{R}}_\downarrow &
        f_\downarrow J'^2 + f_\uparrow D^{\mathrm{R}}_\downarrow D^{\mathrm{A}}_\downarrow
    \end{pmatrix}, \nonumber
\end{align}
where we use the notations
\begin{align}
    & D^{\mathrm{R/A}}_\downarrow(k,\bar{\omega}) = \bar{\omega} +\mu -\tfrac{\Omega}{2} + \epsilon_0(k) \pm i\tfrac{\Gamma}{2} \\
    & D^{\mathrm{R/A}}_\uparrow(k,\bar{\omega}) = \bar{\omega} +\mu +\tfrac{\Omega}{2} -\epsilon_0(k) \pm i\tfrac{\Gamma}{2} \\
    & f_\downarrow(\bar{\omega}) = f(\bar{\omega} -\tfrac{\Omega}{2}) , \quad f_\uparrow(\bar{\omega}) = f(\bar{\omega} +\tfrac{\Omega}{2}).
\end{align}

\subsection{Edge current}
We are now ready to evaluate the physical quantities carried by the electrons on the edge of QSHI.
The current along the edge is given by
\begin{align}
    I &= ie v_{\mathrm{F}} \int_0^\Omega \frac{d\omega}{2\pi} \frac{1}{L} \sum_k \mathrm{Tr}' \left[\sigma_z \mathcal{G}^<(k,\omega)\right],
\end{align}
where $\mathrm{Tr}'$ denotes the trace over the extended Hilbert space $\mathbb{H} \times \mathbb{Z}$.
By evaluating the trace,
we can transform the zone of the $\omega$-integral from a folded zone into an infinite zone,
\begin{align}
    I &= ie v_{\mathrm{F}} \int_0^\Omega \frac{d\omega}{2\pi} \frac{1}{L} \sum_k \\
    & \quad \times \sum_{\nu \in \mathbb{Z}+1/2} \left[ \mathcal{G}^<_{\uparrow\uparrow}(k,\omega+\nu\Omega) - \mathcal{G}^<_{\downarrow\downarrow}(k,\omega+\nu\Omega) \right] \nonumber \\
    &= ie v_{\mathrm{F}} \sum_{\nu} \int_{\nu\Omega}^{(\nu+1)\Omega} \frac{d\bar{\omega}}{2\pi} \frac{1}{L} \sum_k \left[ \mathcal{G}^<_{\uparrow\uparrow}(k,\bar{\omega}) - \mathcal{G}^<_{\downarrow\downarrow}(k,\bar{\omega}) \right] \\
    &= ie v_{\mathrm{F}} \int_{-\infty}^{\infty} \frac{d\bar{\omega}}{2\pi} \frac{1}{L} \sum_k \left[ \mathcal{G}^<_{\uparrow\uparrow}(k,\bar{\omega}) - \mathcal{G}^<_{\downarrow\downarrow}(k,\bar{\omega}) \right] \\
    &= ie v_{\mathrm{F}} \int_{-\infty}^{\infty} \frac{d\bar{\omega}}{2\pi} \frac{1}{L} \sum_k i\Gamma \label{eq:current-int}\\
    & \quad \times \frac{f_\downarrow \left[J'^2 - D^{\mathrm{R}}_\uparrow D^{\mathrm{A}}_\uparrow\right] + f_\uparrow \left[ D^{\mathrm{R}}_\downarrow D^{\mathrm{A}}_\downarrow -J'^2 \right]}{\left[ D^{\mathrm{R}}_{\downarrow} D^{\mathrm{R}}_{\uparrow} - J'^2 \right] \left[ D^{\mathrm{A}}_{\downarrow} D^{\mathrm{A}}_{\uparrow} - J'^2 \right]}. \nonumber
\end{align}

In order to evaluate this integral,
we decompose the integrand into partial fractions.
Here we define the single-band Green's functions
\begin{align}
    g_\pm^{\mathrm{R}}(k,\bar{\omega}) = \left[\bar{\omega} + \mu \mp \xi(k) +i\tfrac{\Gamma}{2}\right]^{-1}, \ g_\pm^{\mathrm{A}} = [g_\pm^{\mathrm{R}}]^*,
\end{align}
with
\begin{align}
    \epsilon(k) &= v_{\mathrm{F}} k +J\cos\alpha -\tfrac{\Omega}{2}, \\
    \xi(k) &= \sqrt{\epsilon^2(k) + J'^2}.
\end{align}
In terms of these single-band Green's functions,
the integrand can be decomposed as
\begin{align}
    \frac{|D^{\mathrm{R}}_\uparrow|^2 - J'^2}{\left| D^{\mathrm{R}}_{\downarrow} D^{\mathrm{R}}_{\uparrow} - J'^2 \right|^2} &= \frac{i}{2\Gamma} \sum_\pm \left[1 \mp \frac{\epsilon}{\xi} -\frac{J'^2}{\xi(\xi \mp i\tfrac{\Gamma}{2})}\right] g_\pm^{\mathrm{R}} + \mathrm{c.c.} \\
    \frac{|D^{\mathrm{R}}_\downarrow|^2 - J'^2}{\left| D^{\mathrm{R}}_{\downarrow} D^{\mathrm{R}}_{\uparrow} - J'^2 \right|^2} &= \frac{i}{2\Gamma} \sum_\pm \left[1 \pm \frac{\epsilon}{\xi} -\frac{J'^2}{\xi(\xi \mp i\tfrac{\Gamma}{2})}\right] g_\pm^{\mathrm{R}} + \mathrm{c.c.}
\end{align}
By using the indefinite integrals
\begin{align}
    \int d\bar{\omega} \ g_\pm^{\mathrm{R}} &= \ln \left[\bar{\omega} +\mu \mp \xi + i\tfrac{\Gamma}{2}\right], \\
    \int d\bar{\omega} \ g_\pm^{\mathrm{R}} g_\pm^{\mathrm{A}} &= \frac{2}{\Gamma} \arctan \frac{\bar{\omega} +\mu \mp \xi}{\Gamma/2},
\end{align}
the $\bar{\omega}$-integrals in Eq.~(\ref{eq:current-int}) can be exactly evaluated as
\begin{align}
    I &= -\frac{e v_{\mathrm{F}}}{2\pi L} \sum_{\nu,\nu' = \pm} \Biggl\{ \frac{-J'^2 \Gamma}{4\xi(\xi^2+\tfrac{\Gamma^2}{4})} \ln\left[ \left(\xi -\nu \tfrac{\Omega}{2} -\nu'\mu\right)^2 +\tfrac{\Gamma^2}{4} \right] \nonumber \\
    & \quad \quad \quad + \left[1-\frac{\nu \epsilon}{\xi} -\frac{J'^2}{\xi^2+\tfrac{\Gamma^2}{4}}\right] \arctan\frac{ \xi -\nu \tfrac{\Omega}{2} -\nu'\mu }{\Gamma/2} \Biggr\} \nu . \label{eq:current-dissipative}
\end{align}
By evaluating the $k$-integral numerically,
we obtain the current $I$ in the presence of the dissipation effect $\Gamma$,
as shown in Fig.~2(a) in the main text.

In the dissipationless limit $\Gamma \rightarrow 0$ with charge neutrality $\mu =0$,
Equation (\ref{eq:current-dissipative}) can be further reduced as
\begin{align}
    I = \frac{e v_\mathrm{F}}{L} \sum_k \left[ \frac{\epsilon}{\xi} \theta(\xi-\tfrac{\Omega}{2}) + \frac{\epsilon^2}{\xi^2} \theta(\tfrac{\Omega}{2}-\xi) \right],
\end{align}
where we have used $\arctan(x/\Gamma) \overset{\Gamma \rightarrow +0}{\rightarrow} \tfrac{\pi}{2} [\theta(x) - \theta(-x)]$ and $2\xi +\Omega >0$.
By evaluating the $k$-integral over $k \in [-k_c,k_c]$ with the cutoff $k_c$,
we obtain
\begin{align}
    I = \frac{e}{2\pi} \left[ \sqrt{\mathcal{E}_{\mathrm{R}}^2+J'^2} - \sqrt{\mathcal{E}_{\mathrm{L}}^2+J'^2} \right] +\frac{e}{\pi}\left[ \mathcal{E}_0 - J' \arctan\frac{\mathcal{E}_0}{J'} \right],
\end{align}
with
\begin{align}
    -\mathcal{E}_{\mathrm{L}} &= -v_\mathrm{F} k_c + J\cos\alpha -\Omega/2 \\
    \mathcal{E}_{\mathrm{R}} &= v_\mathrm{F} k_c + J\cos\alpha -\Omega/2 \\
    \mathcal{E}_0 &= \theta(\tfrac{\Omega}{2}-J') \sqrt{\tfrac{\Omega^2}{4}-J'^2}.
\end{align}
Taking the limit $v_{\mathrm{F}} k_c \gg J, \Omega$,
we obtain the form
\begin{align}
    I &= -\frac{e\Omega}{2\pi} \left[ 1 - \theta(1-\delta)\left( \sqrt{1-\delta^2} - \delta \arctan\sqrt{\delta^{-2}-1} \right) \right].
\end{align}
This is the result shown by the gray solid line in Fig.~2(a).

\subsection{Spin injection rate}
In order to estimate the spin injection rate from the ferromagnet into the QSHI edge,
we need to evaluate the in-plane components of the electron spin accumulation on the edge,
as discussed in the main text.
By using Eq.~(4),
each component can be given in the time-dependent form,
\begin{align}
    \langle \sigma_x(t) \rangle &= -i \int_0^\Omega \frac{d\omega}{2\pi L} \sum_{k,\nu} \left[ \mathcal{G}^<_{\nu,\uparrow\downarrow} e^{-i\Omega t} + \mathcal{G}^<_{\nu,\downarrow\uparrow} e^{i\Omega t} \right] \\
    &= -i \int_{-\infty}^{\infty} \frac{d\bar{\omega}}{2\pi L} \sum_{k} \left[ \bar{\mathcal{G}}^<_{\uparrow\downarrow} e^{-i\Omega t} + \bar{\mathcal{G}}^<_{\downarrow\uparrow} e^{i\Omega t} \right] \\
    &= \int_{-\infty}^{\infty} \frac{d\bar{\omega}}{\pi L} \sum_{k} \left[ \re \bar{\mathcal{G}}^<_{\downarrow\uparrow} \sin \Omega t + \im \bar{\mathcal{G}}^<_{\downarrow\uparrow} \cos \Omega t \right] \\
    \langle \sigma_y(t) \rangle &= -i \int_0^\Omega \frac{d\omega}{2\pi L} \sum_{k,\nu} i\left[ \mathcal{G}^<_{\nu,\uparrow\downarrow} e^{-i\Omega t} - \mathcal{G}^<_{\nu,\downarrow\uparrow} e^{i\Omega t} \right] \\
    &= \int_{-\infty}^{\infty} \frac{d\bar{\omega}}{2\pi L} \sum_{k} \left[ \bar{\mathcal{G}}^<_{\uparrow\downarrow} e^{-i\Omega t} - \bar{\mathcal{G}}^<_{\downarrow\uparrow} e^{i\Omega t} \right] \\
    &= \int_{-\infty}^{\infty} \frac{d\bar{\omega}}{\pi L} \sum_{k} \left[ -\re \bar{\mathcal{G}}^<_{\downarrow\uparrow} \cos\Omega t + \im \bar{\mathcal{G}}^<_{\downarrow\uparrow} \sin \Omega t \right],
\end{align}
where we have used the relation $[\bar{\mathcal{G}}^<_{\downarrow\uparrow}]^* = -\bar{\mathcal{G}}^<_{\uparrow\downarrow}$.
Therefore, the component parallel to $\dot{\boldsymbol{n}}(t) / |\dot{\boldsymbol{n}}(t)| = -\boldsymbol{e}_x \sin\Omega t + \boldsymbol{e}_y \cos\Omega t$ is given as
\begin{align}
    \sigma_{\mathrm{d}} &= -\int_{-\infty}^{\infty} \frac{d\bar{\omega}}{\pi L} \sum_{k}  \re \bar{\mathcal{G}}^<_{\downarrow\uparrow}(k,\bar{\omega}). \label{eq:sigma-d}
\end{align}
Here the integrand reads
\begin{align}
    \re \bar{\mathcal{G}}^<_{\downarrow\uparrow} &= \frac{J'\Gamma^2 \left[f_\uparrow - f_\downarrow \right]}{2 \left[ D^{\mathrm{R}}_{\downarrow} D^{\mathrm{R}}_{\uparrow} - J'^2 \right] \left[ D^{\mathrm{A}}_{\downarrow} D^{\mathrm{A}}_{\uparrow} - J'^2 \right]},
\end{align}
obtained from Eq.~(\ref{eq:G-lesser-matrix}).
By using the decomposition by partial fractions
\begin{align}
    \frac{1}{\left| D^{\mathrm{R}}_{\downarrow} D^{\mathrm{R}}_{\uparrow} - J'^2 \right|^2 } &= \frac{i}{4\xi\Gamma} \sum_\pm \frac{\xi \pm i\tfrac{\Gamma}{2}}{\xi^2 + \tfrac{\Gamma^2}{4}} g_\pm^{\mathrm{R}} +\mathrm{c.c.},
\end{align}
the $\bar{\omega}$-integral in Eq.~(\ref{eq:sigma-d}) can be evaluated at zero temperature as
\begin{align}
    \sigma_{\mathrm{d}} &= -\frac{J'\Gamma}{4\pi L} \sum_k \sum_{\nu,\nu'=\pm} \Biggl\{ \frac{\Gamma}{4\xi} \ln \left[(\xi - \nu\tfrac{\Omega}{2}-\nu'\mu)^2 + \tfrac{\Gamma^2}{4}\right] \nonumber\\
    & \quad \quad + \arctan\frac{\xi-\nu\tfrac{\Omega}{2}-\nu'\mu}{\Gamma/2} \Biggr\} \frac{\nu}{\xi^2 + \tfrac{\Gamma^2}{4}} .
\end{align}
In the regime $J' \gg \Omega, \Gamma$, this quantity gets suppressed.
In particular, this quantity completely vanishes
in the limits $\Gamma \rightarrow 0$, where the heat reservoir is detached from the system,
or $\Omega \rightarrow 0$, where the magnetization is fixed.
Therefore, we should take finite orders in $\Omega$ and $\Gamma$
to evaluate the asymptotic behavior of $\sigma_{\mathrm{d}}$ in the limit $J' \rightarrow \infty$.
At charge neutrality $\mu =0$, the asymptotic behavior becomes
\begin{align}
    \sigma_{\mathrm{d}} &\approx -\frac{J'\Gamma}{2\pi L} \sum_{k,\nu} \frac{\nu}{\xi^2} \left[ \frac{\Gamma}{2\xi} \ln \left( \xi -\nu \frac{\Omega}{2}\right) + \arctan\frac{\xi -\nu \Omega/2}{\Gamma/2} \right] \\
    & \approx -\frac{J'\Gamma}{2\pi L} \sum_{k,\nu} \frac{\nu}{\xi^2} \Biggl[ \frac{\Gamma}{2\xi} \left(\ln \xi -\nu \frac{\Omega/2}{\xi} \right) \\
    & \quad \quad \quad \quad \quad \quad +\arctan \frac{\xi}{\Gamma/2} -\nu \frac{\Omega/\Gamma}{1+(2\xi/\Gamma)^2} \Biggr] \nonumber \\
    & = \frac{J' \Omega \Gamma^2}{2\pi L} \sum_k \frac{1}{\xi^4(k)} \\
    & = \frac{J' \Omega \Gamma^2}{(2\pi)^2} \int_{-k_c}^{k_c} \frac{dk}{\left[(v_{\mathrm{F}}k + J\cos\alpha -\tfrac{\Omega}{2} )^2 + J'^2 \right]^2}.
\end{align}
Since this $k$-integral is free from the ultraviolet divergence,
we can safely extend its zone to $(-\infty,\infty)$ to evaluate its asymptotic behavior.
As a result, the integral reads
\begin{align}
    \sigma_{\mathrm{d}} &\approx \frac{J' \Omega \Gamma^2}{(2\pi)^2 v_{\mathrm{F}}} \int_{-\infty}^{\infty} d\zeta \frac{1}{(\zeta^2+J'^2)^2} = \frac{\Omega \Gamma^2}{8\pi v_{\mathrm{F}}J'^2}. \label{eq:sigma-d-asym}
\end{align}

The spin injection rate can be evaluated by using $\sigma_{\mathrm{d}}$.
The net damping-like torque on the constituent spins in the ferromagnet,
per unit length of the junction, is given by
\begin{align}
    \boldsymbol{T}^{\mathrm{d}}(t) &= J \boldsymbol{\sigma}_{\mathrm{d}}(t) \times \boldsymbol{n}(t) \\
    &= J \sigma_{\mathrm{d}} (-\boldsymbol{e}_x \sin\Omega t + \boldsymbol{e}_y \cos\Omega t) \\
    & \quad \quad \times (\boldsymbol{e}_x \sin\alpha \cos\Omega t + \boldsymbol{e}_y \sin\alpha \sin\Omega t + \boldsymbol{e}_z \cos\alpha ) \nonumber \\
    &= J \sigma_{\mathrm{d}}\left[-\boldsymbol{e}_z \sin\alpha +\cos\alpha(\boldsymbol{e}_x \cos\Omega t + \boldsymbol{e}_y \sin\Omega t) \right].
\end{align}
By taking a time average over a cycle,
the averaged damping-like torque, corresponding to the spin injection rate,
reads
\begin{align}
    -\boldsymbol{J}_{\mathrm{S}} = \overline{\boldsymbol{T}^{\mathrm{d}}} = \int_0^T \frac{dt}{T} \boldsymbol{T}^{\mathrm{d}}(t) =  -J \sigma_{\mathrm{d}} \sin\alpha \ \boldsymbol{e}_z.
\end{align}
By substituting Eq.~(\ref{eq:sigma-d-asym}),
its asymptotic behavior for $J' \rightarrow 0$ is given as
\begin{align}
    \boldsymbol{J}_{\mathrm{S}} \approx \frac{\Omega \Gamma^2}{8\pi v_{\mathrm{F}} J'} \boldsymbol{e}_z.
\end{align}
Thus the asymptotic behavior of the spin-to-charge conversion rate $\lambda_{\mathrm{sc}}$ reaches
\begin{align}
    \lambda_{\mathrm{sc}} &= \frac{I}{-e} \Bigl/ J_{\mathrm{S}}^z \\
    & \approx \frac{\Omega}{2\pi} \Bigl/ \frac{\Omega \Gamma^2}{8\pi v_{\mathrm{F}}J'} \\
    &= \frac{4 v_{\mathrm{F}}J'}{\Gamma^2},
\end{align}
as mentioned in the main text.

\section{Adiabatic pumping picture}
The maximum current $I_c = -e/T$,
which implies that a single electron is pumped along the edge during one cycle of precession,
can be described as the adiabatic (Thouless) pumping.
If the magnetization is static and pointing in the direction 
\begin{align}
    \boldsymbol{n} = \left( \sin\alpha \cos\beta, \ \sin\alpha \sin\beta, \ \cos\alpha \right),
\end{align}
the edge mode opens a gap $2J\sin\alpha$.
The eigenstate wave function in the valence band reads
\begin{align}
    u(k) = \begin{pmatrix} \sin\frac{\theta(k)}{2} e^{-i\beta/2} \\ -\cos\frac{\theta(k)}{2} e^{i\beta/2} \end{pmatrix},
\end{align}
where $\theta(k)$ is the polar angle of the eigenstate spin, defined by
\begin{align}
    \cos\theta(k) = \frac{v_\mathrm{F}k + J\cos\alpha}{\sqrt{(v_\mathrm{F}k + J\cos\alpha)^2 + (J\sin\alpha)^2}}.
\end{align}
If $\beta$ precesses slowly from $0$ to $2\pi$,
we can define the Berry curvature in $(k,t)$-space,
\begin{align}
    \Omega_{kt} = 2 \im \left\langle \frac{\partial u}{\partial t} \Bigl{|} \frac{\partial u}{\partial k} \right\rangle
    = \frac{1}{2} \sin\theta \frac{d\theta}{dk} \frac{d\beta}{dt}.
\end{align}
The number of the electrons pumped per one cycle is given by integrating this Berry curvature over the $(k,t)$-plane, yielding
\begin{align}
    n_{\mathrm{pump}} &= -\int_0^T dt \int \frac{dk}{2\pi} \Omega_{kt} = 1.
\end{align}
This discussion applies as long as an electron cannot be excited by a single magnon with the energy $\Omega$,
from the hole band to the electron band separated by the gap $2J'$.



\begin{thebibliography}{99}
\bibitem{conv-length}
The conversion efficiency $\lambda_{\mathrm{sc}}$ has the dimension of length
due to the mismatch of dimensionalities between $J_C^{\mathrm{int}}$ and $J_S^{\mathrm{bulk}}$.
We denote the charge of an electron by $-e \ (e>0)$ in this article.
\end{thebibliography}
\end{document}